\documentclass[twocolumn,prl,aps]{revtex4}

\def\duzomniejsze{<\kern-.7mm<}
\def\duzowieksze{>\kern-.7mm>}

\def\textbf#1{{\bf #1}}
\def\beq{\begin{equation}}
\def\eeq{\end{equation}}
\def\be{\begin{equation}}
\def\ee{\end{equation}}
\def\ben{\begin{eqnarray}}
\def\een{\end{eqnarray}}
\def\beqa{\begin{eqnarray}}
\def\eeqa{\end{eqnarray}}
\def\eea{\end{array}}
\def\bea{\begin{array}}
\newcommand{\bei}{\begin{itemize}}
\newcommand{\eei}{\end{itemize}}
\newcommand{\bee}{\begin{enumerate}}
\newcommand{\eee}{\end{enumerate}}

\def\tr{{\rm Tr}}
\def\id{{\rm I}}
\def\ra{\rangle}
\def\la{\langle}
\def\>{\rangle}
\def\<{\langle}
\def\blacksquare{\vrule height 4pt width 3pt depth2pt}

\def\ot{\otimes}

\def\logneg{E_{N}}

\newtheorem{lemma}{Lemma}

\newtheorem{proposition}{Proposition}

\newtheorem{definition}{Definition}

\begin{document}

\title{Locking entanglement measures with a single qubit}
 
\begin{abstract}
We study the loss of entanglement of bipartite state subjected to discarding or 
measurement of one qubit. Examining the behavior of  different entanglement measures, 
we find that entanglement of formation, 
entanglement cost, and logarithmic
negativity are {\it lockable} measures in that it can decrease arbitrarily after measuring one qubit. 
We prove that any convex and  asymptotically {\it non}-continuous measure
is {\it lockable}. As a consequence, all the convex roof measures 
can be locked. Relative entropy of entanglement is shown to be a non-lockable measure.
\end{abstract}
\author{ Karol Horodecki$^{(1)}$, Micha\l{} Horodecki$^{(2)}$, 
Pawe\l{} Horodecki$^{(3)}$, Jonathan Oppenheim$^{(4)}$}

\affiliation{$^{(1)}$Department of Mathematics Physics and Computer Science, University of Gda\'nsk, 80--952 Gda\'nsk, Poland}
\affiliation{$^{(2)}$Institute of Theoretical Physics and Astrophysics, University of Gda\'nsk, 80--952 Gda\'nsk, Poland}
\affiliation{$^{(3)}$Faculty of Applied Physics and Mathematics,
Technical University of Gda\'nsk, 80--952 Gda\'nsk, Poland}
\affiliation{$^{(4)}$Dept. of Applied Mathematics and Theoretical Physics, University of Cambridge U.K.}

\maketitle

One of the main differences between quantum and classical information 
are various superadditivities. An example of superadditivity was found 
in \cite{DiVincenzo-locking} where with a single bit, one can lock (unlock)
an arbitrary amount of classical correlations contained in a quantum state (according 
to a physically significant  measure of classical correlations). 
One can ask if similar effects can be found for entanglement. The basic question 
is: how much can entanglement of any bi- or multipartite system change 
when one qubit is discarded? The answer clearly depends on the measure of entanglement.
In this paper we show that the effect of locking holds for the entanglement of formation $E_F$ and 
cost $E_c$, as well as a computable measure of entanglement
- the logarithmic negativity $E_N$ \cite{Vidal-Werner} (cf. \cite{ZyczkowskiHSP-vol}).
 More specifically, we will show that for some state, 
measuring (or dephasing) one qubit can change the entanglement 
from an arbitrary large value to zero. We analyze other entanglement measures. 
We argue that if a measure is convex but not too much, then 
it does not admit locking. We show for example, 
that relative entropy of entanglement can change at most 
by two upon discarding one qubit. Moreover we link the effect of locking 
with the postulate that is often adopted in the asymptotic regime - 
"asymptotic continuity". An entanglement measure is asymptotically continuous, if its 
{\it density} (entanglement per qubit) is continuous, uniformly with respect to dimension. 
The importance of asymptotically continuous measures is that they give rise 
to "macro-parameters" describing entanglement.  I.e. entanglement would be a measure which
changes little if the state changes little.  The effect of locking is a form of 
discontinuity, since by  removing just one qubit, many e-bits 
are destroyed. This raises the question of whether locking is connected to asymptotic continuity. 
We confirm this by proving that a convex measure that 
is not asymptotically continuous admits locking. Our proof is constructive: 
from the states on which a function is discontinuous, one can build a 
state exhibiting locking. Examples are entanglement measures 
built by the convex-roof method \cite{Vidal-mon2000}. 

{\it Entanglement cost and Logarithmic negativity -}  
We shall show that an
{\it arbitrary large} $E_c$, and $\logneg$ of a given state 
can be reduced to zero by a measurement 
on a single qubit. Consider the state on  the Hilbert space 
${\cal H}_{A} \otimes {\cal H}_{B}\sim{\cal C}^{d+2} \otimes {\cal C}^{d+2}$
\be
\rho_{AB}= {1\over 2}\left[
\bea{cccc}
\sigma & 0& 0& {1\over d}U^T \\
0& 0& 0 & 0 \\
0 & 0& 0& 0 \\
{1\over d}(\bar U) & 0& 0& \sigma \\
\eea
\label{blokowa}
\right]
\ee 
Here $U= \sum_{i,j=0}^{d-1} u_{ij} |ii\>\<jj|$ and $\sigma=\sum_{i}\frac{1}{d}|ii\ra\la ii|$ is a separable 
maximally correlated state, and both defined on ${\cal C}^{d}$.  The matrix is written in the computational
basis $|00\ra,|01\ra,|10\ra,|11\ra$ of a pair of qubits each with one of two parties Alice and Bob.  Clearly
after one party measures in the computational basis, the state will decohere, and the off-diagonal elements
will go to zero -- thus the state will be separable.  However, before the measurement, the state has
arbitrarily large entanglement cost i.e. it requires an arbitrarily large number of singlets shared between
Alice and Bob to create, even in the asymptotic limit.  To see this, we take the purification of the state
\ben
\nonumber
\psi_{ABE} = {1\over \sqrt {2d}} \sum_{i=0}^{d-1} \{|i\>|0\>\}_A\{|i\>|0\>\}_B|i\>_E \\
+\{|i\>|1\>\}_A\{|i\>|1\>\}_BU|i\>_E
\een
with the third subsytem denoted as $E$ for Eve (we call the state $\rho_{AE}$ {\it dual} to $\rho_{AB}$).  
One sees that Eve gets a bit string $X$ of length $\log d$ encoded
in one of two basis.  The basis  are complementary if $U$ is taken to be $H^{\otimes \log d}$ with $H$ the Hadamard
transform.
This is precisely the situation for locking classical information
between one party (here AB taken together) and another (here Eve).  
In \cite{DiVincenzo-locking} it was shown that Eve can learn at most $\log d/2$ bits of $X$.  Thus,
for Eve's optimal measurement, the entropy of Alice will be greater than $\log d/2$.  But this is precisely
a definition of $E_F$ i.e.
\beq
E_F(\rho_{AB})=\inf_{A_i}\sum_i p_i S(\rho_i)
\eeq
where the infimum is taken over all measurements with outcomes $A_i$ performed on the purification 
of $\rho_{AB}$ and
resulting in states $\rho_i$ with Alice.  Thus $E_F\geq \log d /2$ which can be arbitrarily large with $d$.  
Furthermore, since the information that Eve can obtain is additive in the number of copies of the state,
$E_c$ is also arbitrarily large.  The log-negativity can also be calculated and it is $\logneg= \log_2(\sqrt{d}+1)$,
thus it too can be locked.

Let us consider another example for locking $\logneg$ which is motivated by the results of \cite{pptkey}. 
To this aim consider the state  defined on the Hilbert space 
${\cal H}_{A}^{(n)} \otimes {\cal H}_{B}^{(n)}$ in such a way that 
${\cal H}_{A}^{(n)}\sim{\cal H}_{B}^{(n)}\sim{\cal C}^{2} \otimes 
({\cal C}^{d})^{\otimes n}$
with natural parameter $n$
and range of parameter $\alpha$ specified subsequently 
\begin{eqnarray}
&&\varrho_{AB}^{(n)}=\frac{1}{2}
[P_{00} \otimes P_{00}\otimes
\tau_{0}^{\otimes n} +  \nonumber \\
&&P_{11}\otimes P_{11} \otimes \tau_{1}^{\otimes n}
+ \alpha^{n}P_{01} \otimes P_{01} \otimes 
(\tau_{1}^{\Gamma}-\tau_{0}^{\Gamma})^{\otimes n}+ \nonumber \\
&&\alpha^{n}P_{10} \otimes P_{10} \otimes  
(\tau_{1}^{\Gamma}-\tau_{0}^{\Gamma})^{\otimes n}].
\label{varrhon}
\end{eqnarray}
Here we use the {\it hiding states} from \cite{WernerHide}
$\tau_{0}=\varrho_{s}^{\otimes l}$, 
$\tau_{1}=({\varrho_{s}+\varrho_{a} \over 2})^{\otimes l}$
where $\varrho_{s}$, $\varrho_{a}$ are fully symmetric 
and antisymmetric Werner states on ${\cal C}^{d} \otimes 
{\cal C}^{d}$.
We also use the notation $P_{ij}=|i\rangle \langle j|$.
The whole matrix can be written 
as before in the form 
\begin{equation}
\varrho_{AB}^{(n)}=\frac{1}{2}
\left[\bea{cccc}\tau_0 ^{\otimes n}&0&0&
\alpha^{n}(\tau_0^{\Gamma}-\tau_1^{\Gamma})^{\otimes n} \\
0&0&0&0\\
0&0&0&0\\
\alpha^{n}(\tau_0^{\Gamma}-\tau_1^{\Gamma})^{\otimes n} 
&0&0&\tau_0^{\otimes n}\\
\eea\right]
\label{raw-key}
\end{equation}
This is a state for any $|\alpha|\leq 1$ since it can be reproduced by 
specific LOCC recurrence protocol \cite{pptkey}
from $\varrho^{(1)}$ defined by the formula above. $\varrho^{(1)}$ 
can be easily checked to be a state. The log-negativity of $\varrho^{(n)}$ 
for given $n$  is
\begin{equation}
\logneg(\varrho^{(n)})=log_{2}[1+(\alpha(2-2^{-l+1})^{n} ] 
\end{equation}
which goes to infinity with $n$ whenever $| \alpha|>(2-2^{-l+1})^{-1}$ 
(this is because of orthogonality of 
$\varrho_{s}$ and $\varrho_{a}$ one has $||\tau_{0}-\tau_{1}||=2-2^{-l+1}$).
On the other hand, measurement of Alice's qubit in the $|i\>$ basis, leads to   
the state $\frac{1}{2} \sum_{i=0}^{1} (|i\rangle\langle i|)^{2}\otimes (\tau_{i})^{n}$ 
which is completely separable.
Hence we have that measurement on a single 
qubit has locked completely an arbitrary high amount of entanglement.

{\it Relative entropy of entanglement.} Let us now examine  relative 
entropy  of entanglement ($E_r$) \cite{VPRK1997}.
We will show that it is not lockable.  More precisely, two solutions will be 
presented, exhibiting that after tracing out one qubit of the 
state $\rho_{AB}$, $E_r(\rho_{AB})$ can decrease at most by two, and 
after a complete von Neumann measurement on one qubit, $E_r$ 
can decrease at most by one.

{\proposition For any bipartite state $\rho_{AA':B}\equiv \rho$  
and any complete von Neumann measurement $\Lambda_{A}$ on the one qbit system $A$ 
there holds:
\be
E_r(\rho)-E_r(\Lambda_{A}\otimes \id_{A'B}(\rho)) \leq 1
\ee
\be
E_r(\rho)-E_r(Tr_{A}(\rho)) \leq 2
\ee
where $Tr_{A}$ denotes partial trace over system $A$.
}

{\bf Proof.}
Both statements of this theorem are consequence of the following property of relative 
entropy of entanglement \cite{LPSW1999} (see \cite{Wilkens} in this context):
\be
{\sum_i p_i E_r(\rho_i)} - E_r({\sum_i p_i \rho_i}) \leq 
S({\sum_i p_i \rho_i}) - {\sum_i p_i S(\rho_i)} 
\label{er}
\ee
where $S$ stands for the von Neumann entropy of the state. 

For the first part of the proof, it suffices to notice that any complete measurement can
be implemented as dephasing of the system. To dephase one qubit, one 
can add a local random ancilla $\tau = {1 \over 2}[|0\>\<0|+|1\>\<1|]$  
and perform the controlled unitary operation $U=\sum_{i=0}^1 |i\>\<i|_{anc}\otimes \sigma^{(i)}_{A}$ with 
$\sigma^{(0)} = \id_{A}$ and $\sigma^{(1)} = \sigma_z$ - a Pauli matrix. 
Indeed, this operation followed by tracing out the ancilla $\tau$ will have the desired effect. 
One can easily check that random unitaries put phases which zero the coherences of the state:
\be
Tr_{anc}[U(\tau \otimes \rho)U^{\dagger}] =
\Lambda_{A}\otimes \id_{A'B}(\rho)\equiv \rho_{meas}
\label{eqstates}
\ee
Taking now in (\ref{er}) $\rho_i = \sigma_i\otimes \id_{A'B}(\rho)$
and $p_i = {1 \over 2}$ one gets 
\be
E_r(\rho)- E_r({\sum_i p_i \rho_i})
\leq
 S({\sum_i p_i \rho_i}) - {\sum_i p_i S(\rho_i)},
\ee
since local unitary transformations do not change $E_r$. 
For such choice of $\rho_i$ and $p_i$ the state ${\sum_i p_i \rho_i}$ is equal to state $\rho$
after dephasing, and by (\ref{eqstates}) is the same as the one after a complete measurement, which gives us:
 
\be
E_r(\rho) - E_r(\rho_{meas})
\leq
S({\sum_i p_i \rho_i}) - {\sum_i p_i S(\rho_i)}.
\label{lasteq}
\ee
It is known \cite{OhyaPetz} that the right hand side does not exceed 
$H(p)$ i.e. the Shannon entropy of the "mixing" distribution $\{p_i\}$. 
In our case this distribution is homogeneous,
so $S({\sum_i p_i \rho_i}) - {\sum_i p_i S(\rho_i)}\leq 1$ which leads us to the first part of the theorem.

The second part of the theorem can be proven in a similar vain. 
Instead of tracing out, we apply total dephasing, which is equivalent to
substitution of a qubit by the maximally mixed one, uncorrelated with the 
rest of the state. To this end we a need bigger random ancilla system $\tau^{\otimes 2}$ 
and the controlled unitary composed from all four Pauli matrices: 
$U=\sum_{i=0}^3 |i\>\<i|_{anc}\otimes \sigma^{(i)}_{A}$. 
The unitaries $\sigma^{(i)}$ are well known examples of ones which when applied randomly change any state to 
the maximally mixed one (see for example, \cite{boykin-qe,mosca-qe}). 

Now the state after the transformation $U$ and tracing out the ancilla $\tau^{\otimes 2}$ is the following: 
$ {\id_A \over 4} \otimes Tr_{A}\rho_{AA'B}$. 
The relative entropy of entanglement of this state is the same as for $Tr_{A}\rho_{AA'B}$,
because it cannot increase after tracing out ${\id_A \over 4}$ 
for this is a local operation, and it cannot decrease, since this qubit is product with the 
rest of the state. 
In this case the right hand side of the inequality (\ref{lasteq}) is bounded by $H(p)=2$ which
completes the proof.

Although it seems to be intuitive, we are not able to show, that both complete measurement
and tracing out of a qubit decrease $E_r$ by the same amount. I.e. for tracing out, we were only
able to prove a bound of $2$ rather than $1$ for the change of $E_r$.  Were this tighter bound to be
proven, one would have an interesting complementarity relation between measuring and
forgetting.  Clearly, measuring a qubit can decrease the entanglement by one ebit.  Likewise,
forgetting the result of a measurement can also decrease entanglement also by one ebit.  An example
of the latter is the measurement result which tells one whether one has a singlet, or some other
Bell state.  Since tracing out a qubit is equivalent to measuring and then forgetting the result,
we would have that if a measurement decreases entanglement by one, then forgetting this result
cannot change the entanglement, and visa versa.

{\it Locking and asymptotic continuity.}
Let us now pass to the connection between asymptotic continuity 
and locking.  Araki and Moriya  \cite{Araki-huge} proved that affine functions are 
Lipschitz continuous. Using similar approach 
Alicki and Fannes \cite{Alicki-Fannes} proved asymptotic continuity of conditional 
entropy which is no longer affine, but is "not too convex". Building on their 
resutls we will prove here  general statement, which can be seen as extension of 
Araki-Moriya result.  Namely, we will  exhibit the following
\begin{proposition} 
\label{prop:affine}
Any function $f$ satisfying 
\bee
\item "approximate affinity": 
$|pf(\rho)+(1-p) f(\sigma)-f(p\rho + (1-p)\sigma)| \leq c$ 
\item "subextensivity" $|f(\rho)|\leq M \log d$
\eee
where $c,M$ are constants, is asymptotically continuous, i.e. it satisfies
\be
|f(\rho_1) - f(\rho_2)| \leq M\|\rho_1-\rho_2\| \log d + 4c
\ee
\end{proposition}
{\bf Remark.} For our purpose (asymptotic regime), it is only important 
that $c$ is constant. However to have also usual continuity, 
it should be that for small $p$, $c$ is small. 
(e.g. for $f$ being  von Neumann entropy,  we have $c\leq H(p)$).

To prove the proposition we need the following lemma of \cite{Araki-huge} 
\begin{lemma}
For any two states $\rho_1\not=\rho_2$, there exist states $\sigma$,
$\gamma_1$ and $\gamma_2$ such that 
\be
\sigma={1 \over (1+\delta)}\rho_1 +{\delta \over (1+\delta)}\gamma_1 
={1 \over (1+\delta)}\rho_2 +{\delta \over (1+\delta)}\gamma_2 
\label{eq:tales}
\ee
where $2\delta=\|\rho_1-\rho_2\|$.
\end{lemma}
{\bf Proof of the Lemma.} One takes states $\gamma_{1(2)}=\omega_\pm/\tr \omega_\pm$, 
where $\omega_\pm$ are positive and negative parts of $\rho_1-\rho_2$. 

{\bf Proof of Proposition.} Let us denote 
$x_i={1 \over (1+\delta)} f(\rho_i) + {\delta\over (1+\delta)} f(\gamma_i) - f(\sigma)$. 
The $x_i$'s 
show how the function $f$ departs from affinity on the considered states.
Positive $x_i$ means convexity, negative $x_i$ means concavity.
Of course $c\geq |x_i|$, because $c$ bounds the departure from affinity 
for any states.  Using (\ref{eq:tales}) we get 
\be
f(\rho_1)-f(\rho_2)= \delta [f(\gamma_1)-f(\gamma_1) ] + (1+\delta)(x_2- x_1)
\label{eq:diff}
\ee
hence due to subextensivity we get 
\be
|f(\rho_1)-f(\rho_2)|\leq 
\delta |f(\gamma_1)-f(\gamma_2)| +(1+\delta)(|x_1|+ |x_2|) \leq \\
2\delta M \log d + 4c \nonumber
\ee
This ends the proof.\blacksquare

Now let us exhibit what happens when a function is subextensive,
but is not asymptotically continuous. To this end 
consider a subextensive function $f$, i.e. let $f(\rho)\leq  M \log d$,
where $\rho$ acts on a $d$ dimensional Hilbert space. Let us assume that $f$ is 
not asymptotically continuous. This means that we have a sequence of states $\rho_1^{(n)}$ 
and $\rho_2^{(n)}$ approaching each other in trace distance,
and acting on a Hilbert space of increasing dimension $d_n$, 
such that 
\be
{|f(\rho_1^{(n)})-f(\rho_2^{(n)})|\over \log d_n}\geq \Delta
\label{eq:divergence}
\ee 
where $\Delta$ is some positive constant. We now consider states  
$\sigma^{(n)},\gamma_1^{(n)},\gamma_2^{(n)}$ given by lemma, 
$\delta^{(n)}={1\over 2}\|\rho_1^{(n)}-\rho_2^{(n)}\|$, and $x_i^{(n)}$ being analogues of $x_i$.
The formula (\ref{eq:diff}) applied to those states together with 
(\ref{eq:divergence}) implies that $|x_1-x_2|\geq (\Delta - 2\delta^{(n)} M) \log d_n$. 
Thus we see that at least one of $x_i$ must have arbitrary large modulus
for large $n$ (i.e. small $\delta^{(n)}$). Without loss of generality, we can assume it is $x_1$.
Then we get that one of two possibilities holds:
\bei
\item[$(i)$] $x_1 \leq (-\Delta/2 +\delta^{(n)} M) \log d_n$
\item[$(ii)$] $x_1\geq (\Delta/2 -\delta^{(n)} M) \log d_n$
\eei
In case $(i)$ the function is too concave, while in case $(ii)$ it is too 
convex. In both cases, the function upon mixing two states 
can be arbitrarily different from the average of the function. 

Let us discuss  the first case. We have a situation 
where upon mixing two states, a function can go up an arbitrary amount.
If $f$ represents e.g. something which is not a valuable resource, then it seems 
not surprising that it can go highly up after forgetting,
as we expect forgetting is not a useful operation. 
If the function is some useful resource, this means that forgetting 
may be very good.  On the other hand, we have the impression that forgetting 
cannot be good for obtaining a resource. Let us explain, 
that the last statement need not be in contradiction 
with an arbitrarily large increase of the function $f$. 
Namely, as noted in \cite{ShorST2001} a function that has such property, 
and is useful  is entanglement of distillation of pure bipartite 
entanglement, from multipartite states. Does it mean that forgetting 
is useful for distillation? It is easily to see that it is not the case (we will 
consider for simplicity two parties).
Simply distillable entanglement of state $\rho$ 
is calculated by taking the product $\rho^{\ot n}$. So $D(\rho)$ represents 
the amount of singlets drawn by Alice and Bob from state $\rho^{\ot n}$ per copy, 
while $D(\sigma)$ the same for state $\sigma^{\ot n}$. 
$D({1\over 2}\rho+{1\over 2} \sigma)$ represents the amount of singlets drawn 
from state $({1\over 2}\rho+{1\over 2} \sigma)^{\ot n}$. We see that 
the latter state cannot be created out of two former states 
by forgetting one bit. The latter would give the much different 
state ${1\over 2}\rho^{\ot n}+{1\over 2}\sigma^{\ot n}$.
Thus for reasonable quantities $f$, the effect $(i)$ should 
be regarded as a type of {\it activation}.

Let us now discuss the case $(ii)$. We have that upon mixing, the function 
goes arbitrarily down. If $f$ is convex, then of course only $(ii)$ 
can occur, and together with convexity, it gives {\it locking}.
We have then the following:
\begin{proposition}
A convex LOCC monotone $E$ that satisfies $E(\rho)\leq M\log d$ 
for some constant $M$, and that is not asymptotically continuous,
admits locking.
\end{proposition}
{\bf Proof.} 
From assumptions it follows that 
there must exist states $\rho_1$ and $\gamma_1$ and weights $1-\epsilon$,
$\epsilon$ such that  the difference 
\be
x=\bigl[\epsilon E(\rho_1) + (1-\epsilon) E(\gamma_1)\bigr] - 
E(\epsilon \rho_1 + (1-\epsilon) \gamma_1)
\label{x}
\ee
can be arbitrarily large. 
Now let us note that a convex entanglement measure  satisfies 
\ben
\nonumber
E(p\rho_{AB}\ot |0\>\<0|_{A'} + (1-p) \tilde\rho_{AB}\ot |1\>\<1|_{A'})=
\\
pE(\rho) + (1-p)E(\tilde\rho) 
\label{add}
\een
One way  follows from convexity and from nonincreasing of $E$ under tracing 
out a local qubit. Second - from the fact that state on the left-hand-side 
of inequality can be transformed into ensemble $\{(p,\rho),(1-p,\tilde\rho)\}$. 
Consider now the state 
\be
\rho_{ABA'}=(1-\epsilon)\rho_1\ot |0\>\<0|_A' + \epsilon \tilde\gamma_1\ot |1\>\<1|_A'
\ee
where $A'$ is one qubit system. Its reduction is given by 
\be
\rho_{AB}=(1-\epsilon)\rho_1 + \epsilon \tilde\gamma_1
\ee
Hence following (\ref{x}) we obtain  that the difference 
\be
E(\rho_{ABA'})-E(\rho_{AB})
\ee
can be arbitrarily large, which is locking. 

{\it Examples.-} Consider so called convex roof measures 
\cite{Vidal-mon2000}, based on Renyi entropy with $0\leq\alpha< 1$.
Such measures are convex by definition, and on pure states 
they are equal to the  Renyi entropy  $S_\alpha= {1\over 1-\alpha} \log \tr \rho^\alpha$ 
of subsystem.
For our choice of $\alpha$ Renyi entropy is greater than von Neumann entropy. 
It is easy to check that for a compressed version of state $\rho^{\ot n}$ 
(denote it by $\rho_{typ}$)
where only typical eigenvalues are kept, the Renyi entropy 
for large $n$ tends to the von Neumann entropy $nS(\rho)$. 
On the other hand for the original state, it is equal to $nS_\alpha(\rho)$.
As we know, the states $\rho_{typ}$ and  $\rho^{\ot n}$ converge to 
each other. However for Renyi entropy we obtain 
that $\Delta= S_\alpha(\rho)-S(\rho)$. Thus Renyi entropy
is not asymptotically continuous, and since we pointed out states 
on which it diverges, one can construct the states, on which 
we have locking effect. 

Let us mention that the above theorem does not say anything 
about measures  which are asymptotically continuous. 
Thus the case of $E_r$ \cite{DonaldH1999} and $E_c$ which are asymptotically continuous 
had to be treated separately.  
Also the theorem does not say anything about measures that are
not subextensive. Therefore the case of negativity was also treated separately. 
We believe that measures such as the distillable entanglement will not be lockable,
but did not prove so here.

Finally we propose a definition of nonlockable version of entanglement measure:
{\definition For any entanglement measure $E(\rho)$ the reduced entanglement measure
$E\downarrow (\rho)$ is defined as
\be
E\downarrow(\rho) = \inf_{\Lambda\in CLOCC}  E(\Lambda(\rho)+ \Delta S)
\ee
}
Here CLOCC is a class of LOCC operations in a closed system and $\Delta S = S(\Lambda(\rho))-S(\rho)$ 
is the increase of entropy produced by measurement. In fact this is quantum analogue of reduced intrinsic information
defined in \cite{renner-wolf-gap}. One can also consider other versions 
of such reduction, choosing maps $\Lambda$ e.g. to be local bistochastic ones or 
local dephasings.

\begin{acknowledgments}
We would like to thank Ryszard Horodecki 
for helpful discussion. This work is supported by 
EU grants RESQ (IST-2001-37559),
QUPRODIS (IST-2001-38877) and
PROSECCO (IST-2001-39227). JO additionally
acknowledges the support of a grant from the Cambridge-MIT Institute.
\end{acknowledgments}

\bibliographystyle{apsrev}
\bibliography{refmich,refjono}

\end{document}